\documentclass[10pt]{iopart}

\input{tcilatex}

\begin{document}

\title{Linear Einstein equations and Kerr-Schild maps}
\author{L\'{a}szl\'{o} \'{A}. Gergely}
\address{
Astronomical Observatory and Department of Experimental Physics,\\ University
of Szeged, D\'{o}m t\'{e}r 9, Szeged, H-6720 Hungary}

\begin{abstract}
We prove that given a solution of the Einstein equations $g_{ab}$ for the
matter field $T_{ab}$, an autoparallel null vector field $l^{a}$ and a
solution $(l_{a}l_{c},\ \mathcal{T}_{ac})$ of the linearized Einstein
equation on the given background, the Kerr-Schild metric $g_{ac}+\lambda
l_{a}l_{c}$ ($\lambda $ arbitrary constant) is an exact solution of the Einstein
equation for the energy-momentum tensor $T_{ac}+\lambda \mathcal{T}
_{ac}+\lambda ^{2}l_{(a}\mathcal{T}_{c)b}l^{b}$. The mixed form of the
Einstein equation for Kerr-Schild metrics with autoparallel null congruence
is also linear. Some more technical conditions hold when the null congruence
is not autoparallel. These results generalize previous theorems for vacuum
due to Xanthopoulos and for flat seed space-time due to G\"{u}rses and G\"{u}rsey.
\end{abstract}

\section{Introduction}

In 1978 Xanthopoulos has proved a remarkable theorem \cite{Xantho},
establishing the linearity of the Einstein equations for an important class
of \textit{vacuum} metrics. According to his result, given an exact solution 
$g$ of the vacuum Einstein equation $R_{ac}[g]=0$ and a null vector field $
l^{a}$, any solution of the linearized vacuum Einstein equation $
dR_{ac}[g(\lambda )]/d\lambda \mid _{\lambda =0}=0$ of the form $
dg_{ac}(\lambda )/d\lambda \mid _{\lambda =0}=l_{a}l_{c}$ corresponds to the 
\textit{exact} vacuum solution $g_{ac}+\lambda l_{a}l_{c}$ ($\lambda $
arbitrary, not necessarily small). Such solutions are of Kerr-Schild (KS)
type \footnote{
Kerr-Schild metrics are frequently considered in the form $\tilde{g}
_{ab}=g_{ab}+V\hat{l}_{a}\hat{l}_{b}$, with $V$ an arbitrary function. Then
a \textit{geodesic curve} with tangent $\hat{l}^{a}$ can be affinely
parametrized such that $\hat{l}^{b}\nabla _{b}\hat{l}^{a}=0$ and the curve
is actually a \textit{geodesic}. We prefer to reparametrize the curve so
that $l_{a}l_{b}$ absorbes the potential $V$, modulo an arbitrary constant $
\lambda $. The price to pay is that affine parametrization cannot be
required for geodesic curves, therefore we call the tangents to geodesic
curves autoparallel vector fields which obey $l^{b}\nabla _{b}l^{a}=fl^{a}$.}
, initially introduced to find exact solutions starting from a flat seed
metric $g_{ac}$ \cite{Trautman,KS}. The theorem proved by Xanthopoulos
assures that exact KS type vacuum solutions can be found by solving the
linearized vacuum Einstein equation.

A systematic study of this linearized equation for vacuum-vacuum KS maps in
a Newman-Penrose framework revealed \cite{GKSlet,GKS1} that the generic case
with the congruence $l^{a}$ shearing does not contain the shear-free case as
a smooth limit. For flat seed space-time the solution of the KS problem is
contained in the Kerr theorem \cite{KerrTh,KerrTh2}, which gives all
shear-free geodesic null congruences $l^{a}$ explicitly. The same theorem
provides valuable help in the search of KS space-times with conformally flat
seed space-times \cite{MaSe}-\cite{So}. The generic shearing case for vacuum
was solved both for twisting \cite{GKS2} and nontwisting null congruences 
\cite{GKSB0}.

An other result relating the linearity of the Einstein equation to KS maps
was established by G\"{u}rses and G\"{u}rsey \cite{GG}. They have shown the
linearity of the \textit{mixed} form of Einstein equations for generic KS
metrics with \textit{flat} seed space-time and \textit{geodesic} KS
congruence. Their proof is based on the remark that for KS space-times with
flat seed metric a Lorentz-covariant coordinate system can be chosen for
which both the Einstein and Landau energy-momentum pseudotensors vanish.
Their result is in a sense more restrictive (flat seed metric, geodesic KS
congruence), in other sense more general (the KS metric not necessarily
vacuum) than the Xanthopoulos theorem. The linearity proved by G\"{u}rses
and G\"{u}rsey however holds only for the mixed form of the Einstein
equation.

The natural question arises whether these two results can be generalized? In
this paper first we find a theorem similar to the one proved by Xanthopoulos
holding in the presence of matter. Basically we establish the set of
conditions to impose on the difference of the seed and KS energy-momentum 
tensors which assure the linearity of the Einstein equation
\footnote{
Our approach differs from previous generalizations allowing the same seed 
and KS energy-momentum tensors which rely on imposition of matter-specific 
constraints on the null KS vector 
\cite{Cast}-\cite{Fuent}.
}. 
For this purpose, in section 2 we derive
the KS equations, which split into a homogeneous and an inhomogeneous
subsystem. With the matter source of the seed metric left arbitrary we seek
the KS energy-momentum tensor in the form of a sum of the seed
energy-momentum tensor and a polynomial in $\lambda $. In the vacuum-vacuum
subcase the result of Xanthopoulos is easily recovered.

In section 3 we solve the homogeneous equations, then we turn to the
analysis of the inhomogeneous subsystem, composed of the third, second and
first order equations in $\lambda $. Then we replace the third and second
order equations with appropiate conditions on the energy-momentum tensor.
These prove our Theorem 1.

Section 4 deals with the particular case of autoparallel congruences. The
main result of the paper is announced as Theorem 2. Then some consequences
of the linear equation are exploited in order to further simplify the KS
energy-momentum tensor.

Finally in section 5 we study the mixed form of the Einstein equation for KS
metrics. In case of generic congruences we find that the Einstein equation
is quadratic. A \textit{sufficient condition} assuring linearity is the
autoparallelism of the congruence. The result of G\"{u}rses and G\"{u}rsey
arises therefore as a special case. We establish and generalize their result
in a covariant manner. Their restriction to flat seed space-time is
unnecesary.

\section{Generalized KS maps}

Two connections $\nabla $ and $\tilde{\nabla}$ annihilating the metrics $
g_{ac}$ and $\tilde{g}_{ac}$ are related \cite{Wald} as: 
\begin{equation}
\tilde{\nabla}_{a}\omega _{b}-\nabla _{a}\omega _{b}=C_{\ ab}^{c}\omega _{c}
\end{equation}
where $\omega _{a}$ is any one-form and $C_{\ ab}^{c}$ is given by 
\begin{equation}
C_{\ ab}^{c}=\frac{1}{2}\tilde{g}^{cd}(\nabla _{a}\tilde{g}_{bd}+\nabla _{b}
\tilde{g}_{ad}-\nabla _{d}\tilde{g}_{ab})\ .
\end{equation}
The relation between the Ricci tensors is straightforward: 
\begin{equation}
\tilde{R}_{ac}=R_{ac}-2\nabla _{\lbrack a}C_{\ b]c}^{b}+2C_{\ c[a}^{e}C_{\
b]e}^{b}\ .  \label{Ricci}
\end{equation}
When the two metrics are the seed metric and the KS metric of the KS map,
defined as 
\begin{equation}
\tilde{g}_{ac}=g_{ac}+\lambda l_{a}l_{c}\ ,\qquad \tilde{g}
^{ac}=g^{ac}-\lambda l^{a}l^{c}\ ,\qquad l_{a}l^{a}=0\ ,  \label{KSmap}
\end{equation}
($\lambda $ arbitrary parameter), Eq. (\ref{Ricci}) takes the detailed form: 
\begin{equation}
\tilde{R}_{ac}=R_{ac}+\lambda R_{ac}^{(1)}+\lambda ^{2}R_{ac}^{(2)}+\lambda
^{3}R_{ac}^{(3)}  \label{Ric}
\end{equation}
with the coefficients $R_{ac}^{(i)}$ given by 
\begin{eqnarray}
R_{ac}^{(1)} &=&\nabla _{b}[\nabla _{(a}(l_{c)}l^{b})-\frac{1}{2}\nabla
^{b}(l_{a}l_{c})]\ ,  \label{Ric1} \\
R_{ac}^{(2)} &=&\nabla _{b}l^{b}l_{(a}Dl_{c)}\!+\!\frac{1}{2}
Dl_{a}Dl_{c}\!+\!l_{(a}DDl_{c)}\!  \nonumber \\
&&+\!l_{a}l_{c}\nabla _{b}l_{d}\nabla ^{\lbrack b}l^{d]}\!-\!Dl^{b}\nabla
_{b}l_{(a}l_{c)}\ ,  \label{Ric2} \\
R_{ac}^{(3)} &=&-\frac{1}{2}l_{a}l_{c}Dl^{b}Dl_{b}\ .  \label{Ric3}
\end{eqnarray}
(We have denoted $D=l^{b}\nabla _{b}$. Round and square brackets on index
pairs denote symmetrization and antisymmetrization, respectively.) For later
convenience we enlist the relations: 
\begin{eqnarray}
R_{ac}^{(1)}g^{ac} &=&(\nabla _{b}D+D\nabla _{b})l^{b}+(\nabla
_{b}l^{b})^{2}\ ,  \label{p1} \\
R_{ac}^{(1)}l^{c} &=&\frac{1}{2}[-Dl^{b}\nabla _{a}l_{b}+\nabla
_{b}l^{b}Dl_{a}+DDl_{a}  \nonumber \\
&&+l_{a}(2\nabla _{b}l_{d}\nabla ^{\lbrack b}l^{d]}\!+\!\nabla _{b}Dl^{b})]\
,  \label{p2} \\
R_{ac}^{(1)}l^{a}l^{c} &=&-Dl^{b}Dl_{b}\ ,  \label{p3} \\
R_{ac}^{(2)}g^{ac} &=&-\frac{1}{2}Dl^{b}Dl_{b}\ ,  \label{p4} \\
R_{ac}^{(2)}l^{c} &=&-\frac{1}{2}l_{a}Dl^{b}Dl_{b}\ ,  \label{p5} \\
R_{ac}^{(3)}l^{c} &=&0=R_{ac}^{(3)}g^{ac}\ .  \label{p6}
\end{eqnarray}
The Einstein equations for the seed and KS space-times 
\begin{equation}
R_{ac}=k(T_{ac}-\frac{1}{2}g_{ac}T)\ ,\qquad \tilde{R}_{ac}=k(\tilde{T}_{ac}-
\frac{1}{2}\tilde{g}_{ac}\tilde{T})\   \label{Ein}
\end{equation}
(where $T=T_{bd}g^{bd}$ and $\tilde{T}=\tilde{T}_{bd}\tilde{g}^{bd}$),
inserted into Eq. (\ref{Ric}) give the KS equation 
\begin{equation}
k(\tilde{T}_{ac}\!-\!T_{ac})\!-\!\frac{k}{2}g_{ac}(\tilde{T}\!-\!T)\!-\!
\frac{k}{2}\lambda l_{a}l_{c}\tilde{T}\!=\!\lambda R_{ac}^{(1)}\!+\!\lambda
^{2}R_{ac}^{(2)}\!+\!\lambda ^{3}R_{ac}^{(3)}\ .  \label{KS}
\end{equation}
We search for the difference of the seed and KS energy-momentum tensors in
the form of a polynomial in $\lambda $ 
\begin{equation}
\tilde{T}_{ac}=T_{ac}+\lambda \mathcal{T}_{ac}+\lambda ^{2}\sigma
_{ac}+\lambda ^{3}\pi _{ac}+\sum_{n=1}^{p}\lambda ^{3+n}\pi _{ac}^{(n)}\ .
\label{expansion}
\end{equation}
(The limit $p\rightarrow \infty $ is allowed.) The parameter $\lambda $
being arbitrary, the KS equation (\ref{KS}) decouples in several equations
corresponding to different powers of $\lambda $. They form a homogeneous 
\begin{eqnarray}
(\lambda ^{5+p}\frac{k}{2}l_{a}l_{c})\ :\quad &&\pi
_{bd}^{(p)}l^{b}l^{d}\!=\!0\ ,  \label{h1} \\
(\lambda ^{4+p}\frac{k}{2})\ :\quad &&g_{ac}(\pi
_{bd}^{(p)}l^{b}l^{d})+l_{a}l_{c}(\pi _{bd}^{(p-1)}l^{b}l^{d}-\pi
^{(p)})\!=\!0\ ,  \label{h2} \\
(\lambda ^{3+n}k\ ,n=\overline{1,p})\ :\quad &&\pi _{ac}^{(n)}+\frac{1}{2}
g_{ac}(\pi _{bd}^{(n-1)}l^{b}l^{d}-\pi ^{(n)})  \nonumber \\
&&\quad +\frac{1}{2}l_{a}l_{c}(\pi _{bd}^{(n-2)}l^{b}l^{d}-\pi
^{(n-1)})\!=\!0\ ,  \label{h3}
\end{eqnarray}
and an inhomogeneous system 
\begin{eqnarray}
(\lambda ^{3})\ &:&\quad k\left[ \pi _{ac}\!+\!\frac{1}{2}g_{ac}(\sigma
_{bd}l^{b}l^{d}\!-\!\pi )\!+\!\frac{1}{2}l_{a}l_{c}(\mathcal{T}
_{bd}l^{b}l^{d}\!-\!\sigma )\right] \!=\!R_{ac}^{(3)}\ ,  \label{la3} \\
(\lambda ^{2})\ &:&\quad k\left[ \sigma _{ac}\!+\!\frac{1}{2}g_{ac}(\mathcal{
T}_{bd}l^{b}l^{d}\!-\!\sigma )\!+\!\frac{1}{2}l_{a}l_{c}(T_{bd}l^{b}l^{d}\!-
\!\mathcal{T})\right] \!=\!R_{ac}^{(2)}\ ,  \label{la2} \\
(\lambda )\ &:&\quad k\left[ \mathcal{T}_{ac}\!+\!\frac{1}{2}
g_{ac}(T_{bd}l^{b}l^{d}\!-\!\mathcal{T})\!-\!\frac{1}{2}l_{a}l_{c}T\right]
\!=\!R_{ac}^{(1)}\ .  \label{la1}
\end{eqnarray}
The brackets preceeding each equation contain the corresponding powers of $
\lambda $ together with the nonessential multiplying factors, which were
dropped. In Eq. (\ref{h3}) $\pi _{ac}^{(0)}=\pi _{ac}$ and $\pi
_{ac}^{(-1)}=\sigma _{ac}$. The traces $\tau ,\ \sigma ,\ \pi $ and $\pi
^{(n)}$ are formed with the metric $g$.

Eq. (\ref{la1}) is obviously the linearized Einstein equation. It can be
obtained independently by the standard linearization procedure, seeking for
approximate solutions $g_{ac}(\lambda )=g_{ac}+\lambda l_{a}l_{c}$ in the
presence of the energy-momentum tensor $T_{ac}(\lambda )=T_{ac}+\lambda 
\mathcal{T}_{ac}$ (with $\lambda $ small). Our aim is to show under what
circumstances Eq. (\ref{la1}) implies the rest of the KS equations (\ref{h1}
)-(\ref{la2}).

\textbf{Vacuum subcase} For vacuum-vacuum KS maps all coefficients in the
expansion (\ref{expansion}) vanish, thus Eqs. (\ref{h1})-(\ref{h3}) are
trivial and Eqs. (\ref{la3})-(\ref{la1}) reduce to $R_{ac}^{(i)}=0,\ i=
\overline{1,3}.$ The twice contracted equation $R_{ac}^{(1)}l^{a}l^{c}=0$ by
virtue of Eq. (\ref{p3}) carries the same information as $R_{ac}^{(3)}=0$,
namely $Dl^{b}=fl^{b}$. Inserting the condition of autoparallelism in $
R_{ac}^{(i)}=0,\ i=1,2$ by virtue of Eq. (\ref{p2}) we find 
\begin{eqnarray}
0 &=&R_{ac}^{(2)}=l_{a}l_{c}\left[ Df+\frac{f^{2}}{2}+f\nabla
_{b}l^{b}+\nabla _{b}l_{d}\nabla ^{\lbrack b}l^{d]}\right] \ , \\
0 &=&R_{ac}^{(1)}l^{c}=l_{a}\left[ Df+\frac{f^{2}}{2}+f\nabla
_{b}l^{b}+\nabla _{b}l_{d}\nabla ^{\lbrack b}l^{d]}\right] \ .
\end{eqnarray}
Thus $R_{ac}^{(2)}=0$ is consequence of $R_{ac}^{(1)}=0$ and all the
information is encoded in the latter equation. Recalling that $
R_{ac}^{(1)}=0 $ is the linearized Einstein equation, we have recovered the
result of Xanthopoulos.

\section{Solution of the nonlinear KS equations}

Next we study the generic system (\ref{h1})-(\ref{la1}).

\textbf{The Homogenous Subsystem} The contraction of Eq. (\ref{h3}) with $
l^{a}l^{c}$ gives $\pi _{ac}^{(n)}l^{a}l^{c}=0$. Then from the trace of Eq. (
\ref{h3}) we find $\pi ^{(1)}=2\pi _{bd}l^{b}l^{d}$ and $\pi ^{(m)}=0,\ m=
\overline{2,p}$. Thus Eq. (\ref{h3}) gives the following solution: 
\begin{eqnarray}
\pi _{ac}^{(1)} &=&\frac{1}{2}g_{ac}(\pi _{bd}l^{b}l^{d})-\frac{1}{2}
l_{a}l_{c}(\sigma _{bd}l^{b}l^{d}-\pi )  \label{pi1} \\
\pi _{ac}^{(2)} &=&\frac{1}{2}l_{a}l_{c}(\pi _{bd}l^{b}l^{d})  \label{pi2} \\
\pi _{ac}^{(k)} &=&0\ ,\qquad k=\overline{3,p}  \label{pi3}
\end{eqnarray}
Eq. (\ref{pi3}) solves the rest of the homogeneous subsystem, Eqs. (\ref{h1}
) and (\ref{h2}) too.

\textbf{The Inhomogeneous Subsystem} The contraction of Eq. (\ref{la1}) with 
$l^{a}l^{c}$ gives 
\begin{equation}
k\mathcal{T}_{ac}l^{a}l^{c}=-Dl^{b}Dl_{b}\ .  \label{la1ll}
\end{equation}
Then the trace and the $l^{c}$ projection of Eq. (\ref{la2}) imply 
\begin{equation}
k\sigma =(-3/2)Dl^{b}Dl_{b}\ , \\
k\sigma _{ac}l^{c}=(-3/4)l_{a}Dl^{b}Dl_{b}\ .
\end{equation}
Inserting these into the third order equation (\ref{la3}), we find: 
\begin{equation}
k\pi _{ac}=-\frac{3}{4}l_{a}l_{c}Dl^{b}Dl_{b}\ .  \label{cond3}
\end{equation}
As a by-product we find $\pi _{ac}^{(1)}=0=\pi _{ac}^{(2)}$. Therefore 
\textit{whenever the KS energy-momentum tensor has the form} 
\begin{equation}
\tilde{T}_{ac}=T_{ac}+\lambda \mathcal{T}_{ac}+\lambda ^{2}\sigma _{ac}-
\frac{3}{4k}\lambda ^{3}l_{a}l_{c}Dl^{b}Dl_{b}\ ,  \label{T2}
\end{equation}
\textit{only the first and second order KS equations are independent}.

Next we remark that the $l^{c}$ projection of the first order equation (\ref
{la1}): 
\begin{equation}
k\left[ \mathcal{T}_{ab}l^{b}+\frac{1}{2}l_{a}(T_{bd}l^{b}l^{d}-\mathcal{T})
\right] =R_{ab}^{(1)}l^{b}\   \label{la1l}
\end{equation}
simplifies the second order KS equation (\ref{la2}). We obtain: 
\begin{equation}
k\sigma _{ac}=kl_{(a}\mathcal{T}_{c)b}l^{b}-\frac{1}{4}
g_{ac}Dl^{b}Dl_{b}+P_{(ac)}^{(2)}\ ,  \label{cond2}
\end{equation}
where 
\begin{equation}
P_{ac}^{(2)}=R_{ac}^{(2)}-l_{c}R_{ab}^{(1)}l^{b}\ .  \label{PPP2dd}
\end{equation}
From Eqs. (\ref{Ric2}) and (\ref{p2}) we find: 
\begin{eqnarray}
2P_{ac}^{(2)} &=&\nabla _{b}l^{b}l_{a}Dl_{c}+Dl_{a}Dl_{c}+l_{a}DDl_{c} 
\nonumber \\
&&+Dl^{b}\left( l_{c}\nabla _{a}l_{b}-2l_{(a}\nabla _{\mid b\mid
}l_{c)}\right) -l_{a}l_{c}\nabla _{b}Dl^{b}\ .  \label{PPP2ddexpl}
\end{eqnarray}
Thus Eq. (\ref{cond2}), expressing $\sigma _{ac}$ in terms of $\mathcal{T}
_{ac},\ g_{ac}$ and $l^{a}$ is a substitute for the second order equation (
\ref{la2}). \textit{Whenever the KS energy-momentum tensor has the
expression }(\ref{T2})\textit{\ and Eqs. }(\ref{cond2})\textit{\ and }(\ref
{PPP2ddexpl})\textit{\ hold, the only independent KS equation is the linear
one.} We have proved:

\begin{theorem}
\textit{Let $g_{ac}$ be a solution of the Einstein equations for
the matter field $T_{ac}$ and $l^{a}$ a null vector field. Any solution $
(l_{a}l_{c},\ \mathcal{T}_{ac})$ of the linearized Einstein equation on the
background $(g_{ac},\ T_{ac})$ corresponds to an exact solution of
Kerr-Schild type $g_{ac}+\lambda l_{a}l_{c}$ (with $\lambda $ arbitrary) of
the Einstein equation with the energy-momentum tensor given by Eqs. (\ref{T2}
), (\ref{cond2}) and (\ref{PPP2ddexpl}).}
\end{theorem}

\section{Autoparallel Congruences}

A more elegant result holds for autoparallel congruences ($Dl^{b}=fl^{b}$).
A simple check shows 
\begin{equation}
P_{ac}^{(2)}=0\ ,  \label{PPP2auto}
\end{equation}
and the content of Eqs. (\ref{T2}) and (\ref{cond2}) reduces to 
\begin{equation}
\tilde{T}_{ac}=T_{ac}+\lambda \mathcal{T}_{ac}+\lambda ^{2}l_{(a}\mathcal{T}
_{c)b}l^{b}\ .  \label{T1geod}
\end{equation}
We announce

\begin{theorem}
\textit{Let $g_{ac}$ be a solution of the Einstein equations for
the matter field $T_{ac}$ and $l^{a}$ an autoparallel null vector field. Any
solution $(l_{a}l_{c},\ \mathcal{T}_{ac})$ of the linearized Einstein
equation on the background $(g_{ac},\ T_{ac})$ corresponds to an exact
solution of Kerr-Schild type $g_{ac}+\lambda l_{a}l_{c}$ (with $\lambda $
arbitrary) of the Einstein equation with the energy-momentum tensor $
T_{ac}+\lambda \mathcal{T}_{ac}+\lambda ^{2}l_{(a}\mathcal{T}_{c)b}l^{b}$.}
\end{theorem}

Some consequences of the remaining linear equation Eq. (\ref{la1}) are
immediate when the congruence is autoparallel.

\begin{theorem}
The null vector of the KS metric is autoparallel if and only if
either of the following conditions hold: 
\begin{equation}
\mathcal{T}_{ac}l^{a}l^{c}=0\Leftrightarrow (\tilde{T}
_{ac}-T_{ac})l^{a}l^{c}=0\ .  \label{geod1}
\end{equation}
\end{theorem}

\begin{proof}
The first statement is obvious from Eq. (\ref{la1ll}). Then $Dl^{b}=fl^{b}$
and (\ref{T1geod}) give the second statement.
\end{proof}

Conversely, Eqs. (\ref{T1geod}) and (\ref{geod1}) imply the autoparallelism
of the null congruence. Our Theorem 3 generalizes Theorem 28.1 of Kramer 
\textit{et al.} \cite{Kramer}, given there for flat seed space-time.

\begin{theorem}
The autoparallel null vector of the KS metric is an eigenvector
of the difference of KS and seed energy-momentum tensors.
\begin{equation}
\mathcal{T}_{ac}l^{c}=\alpha l_{a}\Leftrightarrow (\tilde{T}
_{ac}-T_{ac})l^{c}=\alpha l_{a}\ .  \label{eig}
\end{equation}
\end{theorem}

\begin{proof}
For autoparallel KS congruence the $l^{c}$ projection of the linear
equation, Eq. (\ref{la1l}) becomes 
\begin{equation}
k\left[ \mathcal{T}_{ac}l^{c}\!+\!\frac{1}{2}l_{a}(T_{bd}l^{b}l^{d}\!-\!
\mathcal{T})\right] \!=\!l_{a}\left( Df\!+\!\frac{f^{2}}{2}\!+\!f\nabla
_{b}l^{b}\!+\!\nabla _{b}l_{d}\nabla ^{\lbrack b}l^{d]}\right) \ ,
\label{Tl}
\end{equation}
which implies the first statement. Then the second statement is consequence
of Eq. (\ref{T1geod}) 
\end{proof}

Eq. (\ref{eig}), first derived by Martin and Senovilla \cite{MaSe} was
exploited in the search of KS type perfect fluid solutions \cite{MaSe}-\cite
{So}. The restriction of Theorem 4 to flat seed metric is Theorem 28.2 of
Kramer \textit{et al.} \cite{Kramer}, interpreted as a constitutive
constraint for elastic solid type KS solutions \cite{Magli,EMagli}. Due to
Eq. (\ref{eig}) the energy-momentum tensor (\ref{T1geod}) becomes 
\begin{equation}
\tilde{T}_{ac}=T_{ac}+\lambda \mathcal{T}_{ac}+\lambda ^{2}\alpha
l_{a}l_{c}\ .  \label{T1geod1}
\end{equation}
As consequence, the mixed form of the KS energy-momentum tensor is\textit{\
linear} in $\lambda $: 
\begin{eqnarray}
\tilde{T}_{a}^{\ b}=\tilde{T}_{ac}\tilde{g}^{cb} &=&(g^{cb}-\lambda
l^{c}l^{b})(T_{ac}+\lambda \mathcal{T}_{ac}+\lambda ^{2}\alpha l_{a}l_{c}) 
\nonumber \\
&=&T_{a}^{\ b}+\lambda (\mathcal{T}_{a}^{\ b}-l^{b}T_{ac}l^{c})\ .
\label{Tmix}
\end{eqnarray}

We compute the undetermined coefficient $\alpha $ as follows. Eqs. (\ref{Tl}
) and (\ref{eig}) imply 
\begin{equation}
k\alpha =Df+\frac{f^{2}}{2}+f\nabla _{b}l^{b}+\nabla _{b}l_{d}\nabla
^{\lbrack b}l^{d]}-\frac{k}{2}(T_{bd}l^{b}l^{d}-\mathcal{T})\ .  \label{*1}
\end{equation}
The trace of the linear KS equation (\ref{la1}) gives 
\begin{equation}
k(3T_{bd}l^{b}l^{d}-\mathcal{T)}=2Df+2f\nabla _{b}l^{b}-\nabla
_{b}l_{d}\nabla ^{d}l^{b}+(\nabla _{b}l^{b})^{2}\ .  \label{*2}
\end{equation}
In the derivation of Eq. (\ref{*2}) we have employed the identity 
\begin{equation}
D\nabla _{b}l^{b}=\nabla _{b}Dl^{b}-\nabla _{b}l_{d}\nabla
^{d}l^{b}-R_{bd}l^{b}l^{d}\ ,
\end{equation}
the autoparallelism of $l^{c}$ and the Einstein equations (\ref{Ein}). By
combining Eqs. (\ref{*1}) and (\ref{*2}) we express $\alpha $ in terms of $
f,\ T_{ac},\ g_{ac}$ and $l^{a}$: 
\begin{equation}
2k\alpha =f^{2}+\nabla _{b}l_{d}\nabla ^{b}l^{d}-(\nabla
_{b}l^{b})^{2}+2kT_{bd}l^{b}l^{d}\ .  \label{alpha1}
\end{equation}
Then in the announcement of Theorem 2 one can replace the expression (\ref
{T1geod}) with the set (\ref{T1geod1}) and (\ref{alpha1}).

\section{The linearity of the Einstein equations in the mixed form}

We would like to see, how the previous results relate to the theorem of 
G\"{u}rses and G\"{u}rsey, according to which for flat seed space-time the
mixed form of the Einstein equation is linear in $\lambda $.

We study generic KS congruences. From the KS condition on the metric (\ref
{KSmap}) we find $\tilde{g}_{a}^{b}=g_{a}^{b}=\delta _{a}^{b}$. From Eq. 
(\ref{Ricci}) the following relation between the mixed forms of the Ricci
tensors emerges
\begin{equation}
\tilde{R}_{a}^{b}\!=\!R_{a}^{b}+\lambda P_{a}^{(1)b}\!+\!\lambda
^{2}P_{a}^{(2)b}\!+\!\lambda ^{3}P_{a}^{(3)b}\!+\!\lambda ^{4}P_{a}^{(4)b}\ ,
\label{Riccimix}
\end{equation}
with the coefficients $P_{a}^{(i)b}$ on the right hand side given by 
\begin{eqnarray}
P_{a}^{(1)b} &=&R_{a}^{(1)b}-l^{b}R_{ac}l^{c}\ ,  \label{PP1} \\
P_{a}^{(2)b} &=&R_{a}^{(2)b}-l^{b}R_{ac}^{(1)}l^{c}\ ,  \label{PP2} \\
P_{a}^{(3)b} &=&R_{a}^{(3)b}-l^{b}R_{ac}^{(2)}l^{c}\ ,  \label{PP3} \\
P_{a}^{(4)b} &=&-l^{b}R_{ac}^{(3)}l^{c}\ .  \label{PP4}
\end{eqnarray}
We have denoted $R_{a}^{(i)b}=R_{ac}^{(i)}g^{cb}$. From Eqs. (\ref{Ric3}), 
(\ref{p5}) and (\ref{p6}) we see immediately that $
P_{a}^{(3)b}=P_{a}^{(4)b}=0 $, therefore $\tilde{R}_{a}^{b}$ is only
quadratic in $\lambda .$ By\ inserting the mixed form of the Einstein
equations (\ref{Ein}) into Eq. (\ref{Riccimix}), we obtain the mixed form of
the KS equation: 
\begin{equation}
k(\tilde{T}_{a}^{\ b}\!-\!T_{a}^{\ b})\!-\!\frac{k}{2}\delta _{a}^{\ b}(
\tilde{T}\!-\!T)\!=\!\lambda P_{a}^{(1)b}\!+\!\lambda ^{2}P_{a}^{(2)b}\!\ ,
\label{KSmix1}
\end{equation}
or after the elimination of traces
\begin{equation}
k(\tilde{T}_{a}^{\ b}\!-\!T_{a}^{\ b})\!=\!\lambda \left[ P_{a}^{(1)b}\!-\!
\frac{1}{2}\delta _{a}^{\ b}P^{(1)}\right] +\!\lambda ^{2}\left[
P_{a}^{(2)b}\!-\!\frac{1}{2}\delta _{a}^{\ b}P^{(2)}\right] \ .
\label{KSmix2}
\end{equation}
It is obvious that the difference of the mixed forms of the seed and KS
energy-momentum tensors is quadratic in $\lambda .$ Let us denote then 
\begin{equation}
\tilde{T}_{a}^{\ b}\!=\!T_{a}^{\ b}+\lambda \mathcal{S}_{a}^{b}+\lambda ^{2}
\mathcal{S}_{a}^{(2)b}\ .  \label{Tmix1}
\end{equation}
Decomposition of Eq. (\ref{KSmix1}) with respect to powers of $\lambda $
gives:
\begin{eqnarray}
k\left( \mathcal{S}_{a}^{b}-\frac{1}{2}\delta _{a}^{\ b}\mathcal{S}\right)
&=&P_{a}^{(1)b}\!\ .  \label{mix1} \\
k\left( \mathcal{S}_{a}^{(2)b}-\frac{1}{2}\delta _{a}^{\ b}\mathcal{S}
^{(2)}\right) &=&P_{a}^{(2)b}\!\ .  \label{mix2}
\end{eqnarray}
As before, the contraction of Eq. (\ref{mix1}) with $l_{b}$ results in an
equation with the same r.h.s. as Eq. (\ref{mix2}):
\begin{equation}
R_{a}^{(2)b}-k\left( l^{b}\mathcal{S}_{a}^{c}l_{c}-\frac{1}{2}l_{a}l^{b}
\mathcal{S}\right) =P_{a}^{(2)b}\!\ .  \label{mix1l}
\end{equation}
Comparison gives the \textit{condition, under which the second order
equation is consequence of the first order one}: 
\begin{equation}
k\left( \mathcal{S}_{a}^{(2)b}-\frac{1}{2}\delta _{a}^{\ b}\mathcal{S}
^{(2)}\right) =R_{a}^{(2)b}-k\left( l^{b}\mathcal{S}_{a}^{c}l_{c}-\frac{1}{2}
l_{a}l^{b}\mathcal{S}\right) \ .  \label{S2mix}
\end{equation}
Thus we have the following

\begin{corollary}
Let $g_{ac}$ be a solution of the Einstein equations for the matter field $
T_{ac}$ and $l^{a}$ a null vector field. Any solution $(l_{a}l^{b},\ 
\mathcal{S}_{a}^{\ b})$ of the linearized Einstein equation (in mixed form)
on the given background corresponds to an exact solution $\delta _{a}^{\
b}+\lambda l_{a}l^{b}$ (with $\lambda $ arbitrary) of the (mixed form)
Einstein equation with the energy-momentum tensor given by Eqs (\ref{Tmix1})
and (\ref{S2mix}).
\end{corollary}

For autoparallel null congruences $P_{a}^{(2)b}=P_{ac}^{(2)}g^{cb}$ vanishes
by virtue of Eq. (\ref{PPP2auto}). Therefore the mixed form of the Einstein
equation Eq. (\ref{KSmix1}) becomes linear in $\lambda $.

\begin{corollary}
Let $g_{ac}$ be a solution of the Einstein equations for the matter field $
T_{ac}$ and $l^{a}$ an autoparallel null vector field. Then the mixed form
of the Einstein equation for the KS metric $\delta _{a}^{\ b}+\lambda
l_{a}l^{b}$ and energy-momentum tensor $\tilde{T}_{a}^{\ b}\!=\!T_{a}^{\
b}+\lambda \mathcal{S}_{a}^{b}$ is linear in $\lambda $.
\end{corollary}

Here $\mathcal{S}_{a}^{b}=\mathcal{T}_{a}^{\ b}-l^{b}T_{ac}l^{c}$, cf. Eq. (\ref{Tmix}). The result of G\"{u}rses and G\"{u}rsey, established for flat
seed metric, is contained as a special case.

\section{Concluding Remarks}

The main result of the paper is the generalization of the Xanthopoulos
theorem. We have found the form of the KS energy-momentum tensor for which
the Einstein equations are linear, regardless of the seed energy-momentum
tensor. The importance of both versions of our theorem is given by the
expectation that solving the linearized Einstein equation is easier
(although not trivial at all, as shown by previous analysis of the vacuum
case \cite{GKS1},\cite{GKS2} and \cite{GKSB0}). Still, in the case of
autoparallel null congruences the simplicity of the KS energy-momentum
tensor linearizing the Einstein equation is promising.

The second Corollary generalizes the result of G\"{u}rses and G\"{u}rsey 
\cite{GG} for non-flat seed metrics. Our result relies only on the
autoparallelism of the KS congruence. The equivalent condition on the
energy-momentum tensors is given by Eq. (\ref{geod1}).

Finally we remark that additional constraints on the energy-momentum tensors
possibly emerge, when in order to assure that the local energy density
measured by any local observer is positive, we impose the weak energy
condition on both energy-momentum tensors $T_{ac}$ and $\tilde{T}_{ac}$: 
\begin{equation}
T_{ac}t^{a}t^{b}\geq 0\ ,\qquad \tilde{T}_{ac}\tilde{t}^{a}\tilde{t}^{b}\geq
0\ .  \label{weak}
\end{equation}
Here $t^{a}$ and $\tilde{t}^{a}$ are arbitrary time-like vector fields in
the metrics $g_{ac}$ and $\tilde{g}_{ac}$, respectively. These conditions
are not immediately exploitable in the general case. In the particular case of
algebraically special vacuum seed space-time and autoparallel null
congruence for instance Nahmad-Achar \cite{NA} has obtained from Eq. (\ref
{weak}) a (first order differential) inequality on the spin coefficients.

\ack This work has been completed under the support of the Zolt\'{a}n
Magyary Fellowship. The author is grateful for the financial support of the
organizers of the 9th Canadian Conference on General Relativity and
Relativistic Astrophysics, Edmonton, 2001, where a preliminary version of
this work was presented.

\end{document}